\documentclass[%
 aip,
 jmp,
 amsmath,amssymb,
 preprint,%
]{revtex4-1}
\usepackage{CJK}

\usepackage[colorlinks]{hyperref}
\usepackage{ulem}
\usepackage[T1]{fontenc}
\usepackage{calligra}

\usepackage{graphicx}
\usepackage{dcolumn}
\usepackage{bm}

\usepackage{times}
\usepackage{verbatim}
\usepackage{graphicx}
\usepackage{graphics,epsfig}

\usepackage{theorem}
\usepackage{makeidx}
\usepackage{amsmath}
\usepackage{epic}
\usepackage{amscd}
\usepackage[dvips]{color}
\usepackage{siunitx}
\usepackage{lipsum}
\usepackage{mathrsfs}
\usepackage{bm}

\usepackage{bbm}
\usepackage{xy}
\usepackage{amssymb}
\usepackage{xcolor}
\usepackage{cancel}

\begin{document}

\title{
Statistical mechanics of $d$-dimensional flows and cylindrically reduced passive scalars}
\author{Jian-Zhou Zhu}
\email{jz@sccfis.org}
\affiliation{
Su-Cheng Centre for Fundamental and Interdisciplinary Sciences, Gaochun, Nanjing, 211316 China
}%

%


\date{\today}

\begin{abstract}
Statistical properties of $d$-dimensional incompressible flows with and without cylindrical reduction are studied, leading to several explanations and conjectures about turbulent flows and passive scalars, such as the de-correlation between the flow and scalar, reduction of passive scalar intermittency in the bottleneck regime, et al. The absolute-equilibrium analyses assure the correctness of a recent numerical result. It is implied that passive scalar(s) in two-dimensional (2D) space can be fundamentally different to those in $d>2$, concerning the correlations with the flow, which is not considered in the celebrated Kraichnan model. The possibility of genuine inverse transfer to large scales of 2D passive scalar energy, together with the advection energy, is indicated. The compressible situation is also briefly remarked in the end, in particular the absence of density in a nontrivial Casimir which, without boundary contribution, also vanishes for $d=4$.
\end{abstract}


\maketitle

\section{Introduction}
The investigation of $d$-dimensional ($d$D) hydrodynamics with  $d>3$ can be dated back to Hill\cite{HillTCPS1885} cited by Truesdell,\cite{Truesdell54KinematicsVorticityBook} as noted by Shashikanth.\cite{ShashikanthJMP12} While, those relative to turbulence are relatively new, with the early discussions of flows related to the effects of dimensionality and making the analogy between turbulence and critical phenomena\cite{NelkinPRA74,FournierFrischRoseJPA78} (those works, and some others such as Ref. \onlinecite{LanottePRL15}, studied the non-integer dimension, for reversion of cascade direction, say, but we are only interested in integer $d$ in this note.) And, very recently, turbulent flows in cyclic boxes, or tori $\mathbb{T}^d$, of spatial dimensions $d =4$ and $5$\cite{SuzukiEApof05,GotohETC4DNS07,YamamotoEApre12} and in four-dimensional (4D) channel\cite{NikitinJFM11} have been studied along with direct numerical simulations. However, closely related to our focus of passive scalar(s) from dimensional upgrade and reduction, it is the work of Nordstr\"{o}m\cite{Nordstrom14} where the fifth dimension was introduced and then reduced by cylinder condition along it, resulting in a scalar field.

The (ab)use of the languages of differential geometry will be avoided as much as possible in the discussions to be accessible to more general audiences, for whom however it is still helpful to provide some general background on relevant ideal flows. 
[Readers who are not interested in such background materials can skip.] And, what is more, because the notion of vorticity 2-form is, for $d>3$,  convenient, if not unavoidable in our $\mathbb{R}^d$ or $\mathbb{T}^d$, for explaining the Casimir functions, depending on the parity of $d$ (Ref. \onlinecite{SerreCRM18} and references therein), and because, accidentally in 2D, the duo of velocity and vorticity govern the main features of the statistical hydrodynamics\cite{Lee52} with even the dual cascade,\cite{K67} with the spacial integral of their squares being rugged (in the sense to be explained below) ideal invariants, we quote from Khesin:\cite{KhesinMMJ12} ``One of Arnold's remarkable and, in my opinion, very unexpected insights [in V. I. Arnold, Sur la g\'{e}om\'{e}trie diff\'{e}rentielle des groupes de Lie de dimension infinie et ses applications \`{a} l'hydrodynamique des fluides parfaits. Ann. Inst. Fourier 16, 316--361 (1966)] was to regard the fluid vorticity field (or the vorticity 2-form) as an element of the dual to the Lie algebra of the fluid velocities, i.e., the algebra of divergence-free vector fields on the flow domain.'' Note that two-dimensional (2D) and 3D hydrodynamics are complicated enough, in the sense of both fundamental fluid mechanics, with respect to their volume-preserving diffeomorphism $SDiff$, and fully developed turbulence, with respect also to their irregular solutions. For example, though there exist arbitrary-number-mode `truncations' of the 2D Euler equation to obtain Hamiltonian integrable systems\cite{MurometzRazboynik90} or $sine$-algebra $\mathfrak{sl}(N)$ `analogs' of the Euler Lie algebra $\mathfrak{sdiff}(\mathbb{T}^2)$ [here and after $\mathbb{T}^d = \{(x_1, x_2, ..., x_d) \ \text{mod} \ 2\pi \}$] as $N \to \infty$ holding also `analogously' $O(N)$ Casimir functions\cite{ZeitlinPhD91}, these have nothing to do with the characterization of each of the coadjoint orbits which is generally believed to be nonintegrable; and, the isotopic knots (equipped with the coinciding Kirillov-Kostant and Marsden-Weinstein symplectic structure) identified with a coadjoint orbit, the notoriously hard problem of classifying knot invariants (see, e.g., Ref. \onlinecite{LiuRiccaJFM15} for a recent theoretical study aiming at fluid knot)  becomes (only!) part of classifying all the Casimir functions for $SDiff(\mathbb{R}^3)$ (e.g., Ref.\onlinecite{ArnoldKhesin98Book} who conjectured that ``there are no new integral invariants either for the Euler equation or for the coadjoint orbits of the diffeomorphism groups.'') However, it makes a lot of sense to go to even higher dimensions, for a unified theoretical treatment and universal insights [see, also, recently, Fecko\cite{FeckoAPS14,FeckoJGP17} for generalization of the Helmholtz theorems for vortex lines, Besse \& Frisch\cite{BesseFrischJFM17} for a geometric account, and also Serre\cite{SerreCRM18} for a review relevant to integrals of motion.]

Only quadratic invariants are known from 2D and 3D Galerkin truncated incompressible Euler systems for absolute equilibrium analyses\cite{Lee52,K67,K73} of turbulence. The $sine$-algebra approach, may have other merits, say, for numerical approximation, has not been more powerful in this analytical line. And, although for some specific systems with non-quadratic invariants survived from the Galerkin truncations (rugged) as nicely documented (e.g., Refs. \onlinecite{AbramovKovacicMajdaCPAM03,KrstulovicETC09,KrstulovicBrachet11}), it is hard to obtain analytical insights. With Gaussian and adiabatic approximations under favorable conditions, as Kraichnan\cite{K55} did, but additionally respecting the role of helicity, useful physical predictions such as the reduction of the compressible modes (thus presumably the turbulent aeroacoustic noise) of a gas can be made.\cite{ZhuJFM16} 
Our main point is that some insights about the statistical mechanics of 2D and 3D (multiple) passive scalar(s) can be obtained by combining the techniques of dimensional upgrade and reduction (by cylinder conditions) with the absolute equilibrium analysis. 

Passive scalar advected by an ideal incompressible flow just labels the $SDiff$ with no extra information. Intriguing things such as (turbulence-enhanced) mixing is related to the irregular fluid advection and molecular diffusion.\cite{VillermauxARFM19}
Just as the fluid viscosity, the damping effect is stronger (faster) for higher harmonic modes for the scalar diffusion, which ``implies that a phase point of the infinite-dimensional space is attracted to the finite-dimensional one, where the coordinates are the amplitudes of the lower harmonics''.\cite{ArnoldKhesin98Book} Thus, Galerkin truncation of the pure advection of a passive scalar is accordingly related to a diffused scalar. Keeping the aforementioned in mind, it is actually in turn possible to extract reasonable information about real turbulence from the truncated advection. For example, attentions have been paid to the phenomena of de-correlation between the passive scalar and the flow, and, between multiple scalars\cite{sirivatWarhaftJFM82,YeungPopePoF93,JunejaPopePoF96,YeungPoF98,GotohYeung10} (everything ultimately becomes uniform and fully correlated in the damping case,\cite{VillermauxARFM19} but we are interested in the regimes dominated by nonlinear interactions.) However, to our best knowledge, no guiding principle from the dynamics has been established. It is intriguing whether the corresponding ideal invariants for any $d$ other than $2$ and $3$\cite{ArnoldKhesin98Book} has  genuine effects on the passive scalar(s) from the dimensional reduction with \textit{cylinder condition}. A relevant explanation about cylindrical reduction is offered in Appendix \ref{apd:cylinder}, and ruggedness of invariants is elaborated in Appendix \ref{apd:ruggedness} for readers in need.

\section{Basic equations}
To be more definite, let's write down the passive-scalar Navier-Stokes (NS) equations in $d$D
\begin{eqnarray}
  \partial_t \theta + \bm{v}\cdot\nabla\theta&=& \kappa\nabla^2\theta+f_{\theta},\label{eq:p1} \\
  \partial_t \bm{v} + \bm{v}\cdot \nabla \bm{v}&=& -\nabla P+\nu \nabla^2\bm{v}+\mathbf{f}_{\bm{v}}, \label{eq:p3}\\
  \nabla \cdot \bm{v}&=&0,   \label{eq:incompressibility}
\end{eqnarray}
The scalar $\theta$ is \textit{passive} with no back-reaction onto $\bm{v}$ through the forcing $\mathbf{f}_{\bm{v}}$ which is independent of $\theta$ (otherwise \textit{active}). $\bm{v}$ is the incompressible NS velocity field (when $\mathbf{f}_{\bm{v}}$ is not a function of $\bm{v}$), and the pressure $P$ satisfies the Poisson equation from taking the divergence of (\ref{eq:p3}). 
Curling (\ref{eq:p3}) for $d=2$, we have
\begin{equation}\label{eq:p2}
  \partial_t \zeta + \bm{v}\cdot\nabla\zeta = \nu\nabla^2\zeta+ \nabla \times f_{\bm{v}} \cdot \bm{z},
\end{equation}
with the vertical vorticity
$\zeta \bm{z}=\bm{\zeta}=\nabla \times \bm{v}.$
For the incompressible NS equation with $\partial_z \equiv 0,$
i.e., depending only on $x$ and $y$ coordinates, or averaged over $z$, the velocity $\bm{u}$ denotes a 2D3C flow, and the vertical velocity $\bm{u}_z=u_z \bm{z}=\theta\bm{z}$ in the (unit) $\bm{z}$ direction is the third component passively advected by the horizontal velocity $\bm{u}_h=\bm{v}$, thus a problem of 2D passive scalar with
$\nu/\kappa=1.$ Reversely, one can `upgrade' the 3D passive scalar problem to be of the 4D NS dynamic with cylinder condition, and with component-dependent viscosities if the Prandtl/Schmidt number is non-unit. And, we can have multiple passive scalars through multiple such cylinder reductions of dimensions. 

The dynamical invariants or the constants of motion present reduced forms in accordance with the cylinder reductions. In addition to the kinetic energy, others have been found to be conserved by $d$-dimensional ($d$D) ideal flows by L. Tartar ($\mathscr{T}$) and Serre ($\mathscr{S}$), respectively for odd and even $d$ (Ref. \onlinecite{SerreCRM18} and references therein). And, it is interesting to note that the $d$D($d$+1)C system brings all $d$D and $(d+1)$D invariants altogether, and especially both $\mathscr{T}$ and $\mathscr{D}$ are held simultaneously from such dimensional reduction. For instance, the 2D3C global helicity turns into the cross-correlation between the 2D vorticity and the third component, $\langle \zeta \theta \rangle$, besides the kinetic energy and enstrophy of the horizontal flow, the total kinetic energy, which should be respected in relevant statistical analyses.\cite{ZhuPoF18Schur}

In 4D, it is convenient to switch the notation convention: $$\text{$x \to 1$, $y \to 2$, $z \to 3$ and the fourth velocity component $u_4 = \theta$}.$$ The latter is a passive scalar with the dynamics (\ref{eq:p1}, \ref{eq:p3}) when the above cylinder condition is applied with respect to its fourth coordinate. And, in contrast to Eq. (\ref{eq:p2}), we have
\begin{equation}\label{eq:p4}
\partial_t \bm{\omega} = \nabla \times (\bm{v} \times \bm{\omega}) + \nu \nabla^2 \bm{\omega} + \nabla \times \bm{f}_{\bm{v}},
\end{equation}
with $\bm{\omega} = \nabla \times \bm{v}$ being the the vorticity of the 3D horizontal velocity $\bm{v}=\bm{u}_h = (u_1, u_2, u_3)$. However, as we will see, the reduction of a quadratic $\mathscr{T}$ in 3D4C does not simply result in a correlation as in 2D3C: The relevant quantity is the cross-correlation between the gradient of the fourth velocity component $\theta$ and the 3D vorticity $\bm{\omega}$,
\begin{equation}\label{eq:3D4Chelicity}
\text{$\mathcal{N} = \langle \nabla \theta \cdot \bm{\omega} \rangle$, which is null when the boundary, if exists, contributes nothing extra.}
\end{equation}
The general issue of the effects of $d$D-invariant constraint after dimensional reduction, especially on the passive scalar(s), is intriguing.

Although it has been around seventy years since Onsager\cite{Onsager49} argued ```ideal' turbulence'' for the inviscid limit of the 3D Euler dynamics, few systematic methods exist to effectively shed light on fundamental issues such as the directions of the spectral transfers;\cite{Frisch95book,EyinkPhD08} and, use will be made of a tool of statistical analysis called absolute equilibrium of the Galerkin-truncated inviscid system, indicated by Onsager\cite{OnsagerLetter45} himself and with others' independent discoveries and developments.\cite{Lee52,K67} It is interesting to note that in the early days, like Lee,\cite{Lee52} Ziman ,\cite{Ziman53} in the context of quantum hydrodynamics as the theoretical efforts for the roton spectrum of the liquid helium, referred the Galerkin truncation of the Fourier modes to the interatomic dimension, and the analogy with Debye's truncation of phonon frequency was pointed out. As discussed in Frisch et al.,\cite{FrischETAprl08} the Galerkin truncation for preparing the absolute equilibria may be regarded as the effect of infinite dissipation rate for the truncated modes, and, the finite-dissipation-rate dynamics inbetween the Euler and Galerkin-truncated one presents partial thermalization, which explains the spectral bottleneck and intermittency growth deceleration, the latter of which was indeed observed in high-resolution direction numerical turbulence of Newtonian fluid\cite{ZhuCPL06} and, more obviously, of high-order dissipative models:\cite{ZhuTaylorCPL10} As will be elucidated and expected, similarly is the case for 3D passive scalars (c.f., Fig. 3.3 of Ref. \onlinecite{GotohYeung10}), but not for 2D.
We however want to put the methodology to the extremes to check more seriously the statistical mechanics of passive scalars and of higher-dimensional flows in a unified way. The value will be proved by explaining/predicting the relevant observations. It is expectable to see further potential usage also for other systems, such as the Gross-Pitaevskii quantum turbulence whose \textit{in silico} 4D realizations have been made recently.\cite{MiyazakiEAphd10}

\section{Dynamical invariants and statistical mechanics analyses}\label{sec:analyses}
Dynamical invariants of the ideal incompressible flows are crucial in the analyses below, so it is important to give an overall remark before delving  into the specific details.
For simplicity, the domain $\mathcal{D}$ of the fluid location $\bm{r}$ is $\mathbb{R}^d$ or $\mathbb{T}^d$ (i.e., a cyclic box, with period $2\pi$ as indicated in the beginning, over which all integrations are performed by default) with Euclidean structure, although some of the formulations also apply to (pseudo-)Riemannian manifold, and correspondingly the Fourier (wavenumber $\bm{k}$) space is considered.
In the current physical consideration only those observables conserved in detail or rugged with Galerkin truncation, thus quadratic in general, are relevant constraints:\cite{Onsager49,Lee52,K67,K73} Kinetic energy always has a seat, and enstrophy and helicity take roles in 2D and 3D respectively.

We need to use some differential-geometry language and notation conventions (the same as in Zhu\cite{ZhuPoF18Lie}), such as $\verb"U"$ for the 1-form velocity (corresponding to $\bm{u}$) and its exterior derivative $\verb"d" \verb"U"$ as a \textit{vorticity (2-)form}. For the ideal Euler flow in a manifold $\mathcal{M}$ (more general than our $\mathcal{D}$) of dimension $d$ the following invariants are known\cite{SerreCRM18}
\begin{eqnarray}
\mathscr{T}=\int_{\mathcal{D}} g(\tau) \mu \  \text{with the \textit{vorticity function}} \ \tau = (\verb"d" \verb"U")^m / \mu \  \text{for even $d=2m$} \label{eq:dDinvariantsT}
\end{eqnarray}
and any smooth function $g$, and, the (generalized) helicity
\begin{eqnarray}
\mathscr{S}=\int_{\mathcal{D}} \verb"U" \wedge (\verb"d" \verb"U")^m \  \text{for odd $d=2m+1$},\label{eq:dDinvariantsD}
\end{eqnarray}
with the power $m$ denoting $m$ times of wedge products, and $\mu$ the volume form. For odd $d$, the \textit{vorticity vector} $\bm{\omega}$ (as the generalized \textit{curl} of $\bm{u}$) satisfies the \textit{i}nterior product relation $i_{\bm{\omega}} \mu = (\verb"d" \verb"U")^m$. We need to identify in the following discussions which of the above invariants are relevant and how they take effect. It is hardly possible there be other quadratic ones when the degrees of freedom (in the sense of Fourier modes below) are large.

The reason why we consider the non-quadratic invariant be generally irrelevant to our absolute equilibrium calculations here can be heuristically explained with a caveat elaborated by the Burgers model, as given in the Appendix \ref{apd:ruggedness}.

So far, systematic knowledge about passive scalar is available only for the Kraichnan\cite{K68} model with independent synthetic velocity and pumping, both of which are delta correlated in time.\cite{FGVrmp01, CCMVnjp04} And, a kind of `unified' theory, with, say, $f_{\theta}$ also possibly depending on $\bm{v}$ in some specific way (relevant to the injection of the correlation of 2D passive scalar or the helicity of the 2D3C flow to be discussed below), is practically wanted for the `self-consistent' dynamics.

\subsection{4D incompressible flows and 3D passive scalar}
For a 4D $\bm{u}$ solving Eq. (\ref{eq:p3}) for $\bm{v}$, $\verb"d" \verb"U"$ reads in co-ordinate form
\begin{eqnarray}\label{eq:4Dcurl}
\verb"d" \verb"U" = (u_{2,1}-u_{1,2}) dx_1 \wedge dx_2 + (u_{3,1}-u_{1,3})dx_1 \wedge dx_3 + (u_{4,1} - u_{1,4})dx_1 \wedge dx_4 +  \nonumber\\
+ (u_{3,2}-u_{2,3})dx_2 \wedge dx_3 + (u_{4,2} - u_{2,4})dx_2 \wedge dx_4 + (u_{4,3} - u_{3,4})dx_3 \wedge dx_4,\\
\text{with} \ \verb"U" = u_1 dx_1 +u_2 dx_2 + u_3 dx_3 + u_4 dx_4.\nonumber
\end{eqnarray}
And we obtain $\verb"d" \verb"U" \wedge \verb"d" \verb"U" = \tau dx_1 \wedge dx_2 \wedge dx_3 \wedge dx_4$, with a factor of two neglected below for convenience, 
\begin{eqnarray}\label{eq:tau}
\tau = && (u_{1,2}-u_{2,1})(u_{3,4}-u_{4,3})-(u_{1,3}-u_{3,1})(u_{2,4}-u_{4,2})+(u_{1,4}-u_{4,1})(u_{2,3}-u_{3,2}) \nonumber \\
=&& (u_{1,2}u_{3,4}-u_{1,4}u_{3,2}) + (u_{1,3}u_{4,2}-u_{1,2}u_{4,3}) + (u_{3,1}u_{2,4}-u_{2,1}u_{3,4}) +  \\
&& + (u_{2,1}u_{4,3}-u_{4,1}u_{2,3}) + (u_{1,4}u_{2,3}-u_{1,3}u_{4,2}) + (u_{3,2}u_{4,1}-u_{3,1}u_{4,2}).\nonumber
\end{eqnarray}
Now $\tau$ is quadratic and, as said, we believe that 
\begin{equation}\label{eq:4Dhelicity}
\mathscr{N} = \int_{\mathcal{D}} \tau \mu
\end{equation}
is the only rugged invariant among all $g$s in Eq. (\ref{eq:dDinvariantsT}). 
And, organising $\tau$ differently in Eq. (\ref{eq:tau}) are essential for the following two important observations respectively:

\textit{First}, we can see from the right hand side of the second equality that, when the boundary contribution from integration by parts [corresponding to the \textit{Stokes theorem} for the integration of differential forms]
is null, as particularly for $\mathcal{D} = \mathbb{T}^4$,
\begin{equation}\label{eq:nullN}
\mathscr{N}=0.
\end{equation}
Shashikanth\cite{ShashikanthJMP12} also observed the corresponding result for $\verb"d" \verb"U"$ decaying sufficiently fast at infinity in $\mathbb{R}^4$. [Actually, since $\verb"d" \verb"U" \wedge \verb"d" \verb"U"=\verb"d" (\verb"U" \wedge \verb"d" \verb"U")$, we can establish the generalized Cauchy invariant equation\cite{BesseFrischJFM17} and also the generalized Helmholtz theorems\cite{FeckoAPS14,FeckoJGP17} to expose more fundamentals of 4D flows, which however is not the interest here and will be communicated elsewhere.] Thus, the only constraint left is the obvious one of kinetic energy which is equipartitioned in the absolute equilibrium state of the Galerkin truncated inviscid system.\cite{Lee52} This result is nontrivially trivial, because, otherwise, dual cascades as in 2D or 3D turbulence would happen; and, it explains the findings of Suzuki et al.\cite{SuzukiEApof05} and offers the answer to the puzzle why the quadratic, thus rugged, invariant $\mathscr{T}$ has no effect. We are now interested in the intrinsic dynamics free from boundary effects, and the absolute-equilibrium equipartition and turbulent forward cascade of energy they found should be genuine and, as will be argued, common to all $d>3$. As a side note, on the other hand, this invariant is a good measure of the boundary effects, which may be useful in other situations, such as the channel flows as studied by Nikitin.\cite{NikitinJFM11}

\textit{Second}, as said, with the `cylinder' condition $\partial_{r_4} = 0$, $u_4= \theta$ becomes a passive scalar, if the forcing on the `horizontal' 3D velocity does not depend on $u_4$. For such a 3D4C system, the rugged invariant constraints are the well-known kinetic energy and helicity of the 3D advecting (`horizontal') $\bm{v}$,\cite{K73} and, the energy of the passive scalar $\mathcal{Z}=\langle \theta^2 \rangle$ (the difference between the total 4D kinetic energy and horizontal 3D one): $\mathscr{N}$ still vanishes by logic, but it is also direct to check term by term in the first line of Eq. (\ref{eq:tau}) that
\begin{equation}\label{eq:3D4CN}
\text{$\tau \to \omega_3 \theta_{,3}+\omega_2 \theta_{,2}+\omega_1 \theta_{,1} = \nabla \theta \cdot \bm{\omega} = \nabla \cdot (\bm{\omega} \theta)$, and that $\mathscr{N} \to \mathcal{N}$ in (\ref{eq:3D4Chelicity})}.
\end{equation}
[$\mathcal{N}=0$ is seen also from the vanishing expression in each spectral component, $\bm{k} \cdot (\bm{k} \times \hat{\bm{v}}_{\bm{k}}) = 0$, the reduced case of the fact that, each pair in the six parentheses, behind the second equality of Eq. (\ref{eq:tau}) for $\tau$, cancels their spectra and of course cancels after spatial integration over the corresponding two directions.] So, $\mathcal{N}$ produces null cross-correlation constraint between $\theta$ and $\bm{v}$. And, the commonly accepted equipartition of absolute equilibrium $\mathcal{Z}$, obtainable from the canonical ensemble, is assured, indicating genuine forward turbulent cascade, since the diffusivity will damp the large-wave-number modes to whom the mode interactions then keep transfer energy, tending to repair the incomplete thermalization.\cite{K67} Similar to 4D wall-bounded flow such as the channel flow,\cite{NikitinJFM11} 3D passive scalar in wall-bounded flow\cite{DrivasEyinkJFM17II} may present the effect of non-zero $\mathcal{N}$.

It is interesting to digress to remark that Khesin and Chekanov\cite{KhesinChekanovPhD89} note that the
reduction with ``shear plane-parallel flows'' ``from even to odd dimensions does not provide any new integrals''.\cite{ArnoldKhesin98Book} Now the cylindrical reduction do provide `new integrals' which however are trivial in the sense of dynamics, since these invariants do not offer any information or constraint on the lower odd dimensional flow but only mean that the extra dimensional velocity component(s) needs to change in a passive way to keep the integrals.

As the residue of the partial thermalization, just as the kinetic energy spectral bottleneck\cite{FrischETAprl08} as mentioned in the introductory discussions, we expect that, if the Prandtl/Schmidt number is unit (thus no other effects due to different dissipation scales), a similar phenomena should present in the spectra of the 3D passive scalar inbetween the inertial and dissipation ranges, which indeed is the case (c.f., Fig. 3.3 of Ref. \onlinecite{GotohYeung10}). Of course, further verification of such an interpretation/prediction should be made with extra measurement of the ``intermittency growth rate''\cite{ZhuCPL06} of the scalar increments, and we further expect that, similarly, ``a lull in the growth of intermittency at bottleneck scales may already be observed'',\cite{FrischETAprl08} and can be more obvious for models of hyperviscosity and hyperdiffusivity fashion, as presented in Zhu and Taylor\cite{ZhuTaylorCPL10} for the flow with hyperviscosity-like model.

\subsection{5D hydrodynamic turbulence and multiple 3D and 2D passive scalars}\label{sec:2D}
It is seen that\cite{ZhuPoF18Lie} for the 2D3C Euler, the (\textit{Lie}-)invariant
$\text{\textit{local helicity}} \ \verb"h"=2\theta \zeta$
may be used as a surrogate for the spatial density $h = \nabla \times \bm{u} \cdot \bm{u}$ of the invariant
$\text{\textit{global helicity} $\mathcal{H} = \int h d\bm{r}$}$
with such appropriate (say, periodic) boundary conditions that no boundary term appears from the integration by parts.
That is, $\mathcal{H}$ reduces to the $\theta$-$\zeta$ cross-correlation (see also Moffatt\cite{MoffattJFM69})
\begin{equation}\label{eq:2D3Chelicity}
\mathcal{C} = \langle 2\theta \zeta \rangle = \int 2\theta \zeta d\bm{r}.
\end{equation}
Thus, the global invariance of helicity can be either Lagrangian with the domain surrounded by the moving boundary $\partial \mathsf{D}(t)$, or  Eulerian with fixed domain $\mathcal{D}$ where the flow needs to satisfy some particular (say, periodic) conditions on $\partial \mathcal{D}$.

Decomposing the 2C2Dcw1C3D field into the horizontal and vertical parts as conventionally done in the 2D3C flow leads to the similar dynamics of the decomposed vorticities and helicities to those of the latter;\cite{ZhuPoF18Schur} and, if periodic boundary conditions were applicable, the same rugged global conservation laws as the following would formally hold: the \textit{(horizontal) kinetic energy}
$\mathcal{E}=\langle u_h^2\rangle
$, the \textit{enstrophy} $\mathcal{W}=\langle\zeta^2\rangle
$, and the \textit{passive-scalar (vertical) energy} $\mathcal{Z}=\langle\theta^2\rangle=\langle u_z^2\rangle$, besides $\mathcal{C}$. The absolute equilibrium spectra for 2D3C flows was calculated from canonical (joint) probability distribution function [(J)PDF] $\sim \exp\{-(\Gamma_{\mathcal{C}} \mathcal{C} + \Gamma_{\mathcal{E}} \mathcal{E} + \Gamma_{\mathcal{W}} \mathcal{W} + \Gamma_{\mathcal{Z}} \mathcal{Z}) \}$, with $\Gamma_{\bullet}$ being the corresponding Lagrangian multipliers (temperature parameters) for the invariants\cite{K67} of the corresponding Galerkin truncated system in a cyclic box (of dimension $L=2\pi$) with the Fourier transform $\bm{v}(\bm{r}) \leftrightarrow \hat{\bm{v}}_{\bm{k}}$ (and similar Fourier representations for other variables, all with vanishing vertical component of the wavenumber $\bm{k}$, $k_z = 0$): with
\begin{eqnarray}
D=\Gamma_{\mathcal{E}}\Gamma_{\mathcal{Z}}+(\Gamma_{\mathcal{W}}\Gamma_{\mathcal{Z}} - \Gamma^2_{\mathcal{C}})k_h^2>0, \ U_h \triangleq \langle |\hat{u}_h|^2 \rangle = \frac{\Gamma_{\mathcal{Z}}}{D}, \ W \triangleq \langle |\hat{\zeta}|^2 \rangle =k_h^2 U_h, && \label{eq:2D3CU}\\
Q_{\mathcal{C}}\triangleq \langle \hat{\zeta}\hat{\theta}^*\rangle+c.c.= \frac{-2\Gamma_{\mathcal{C}}k_h^2}{D} \ \text{and} \ U_z \triangleq 
\langle |\hat{\theta}|^2 \rangle= 
\frac{\Gamma_{\mathcal{E}}+\Gamma_{\mathcal{W}}k_h^2}{D}. &&\label{eq:2D3CQ}
\end{eqnarray}

\begin{figure}
  \centering
  \includegraphics[width=
  \paperwidth
  ]{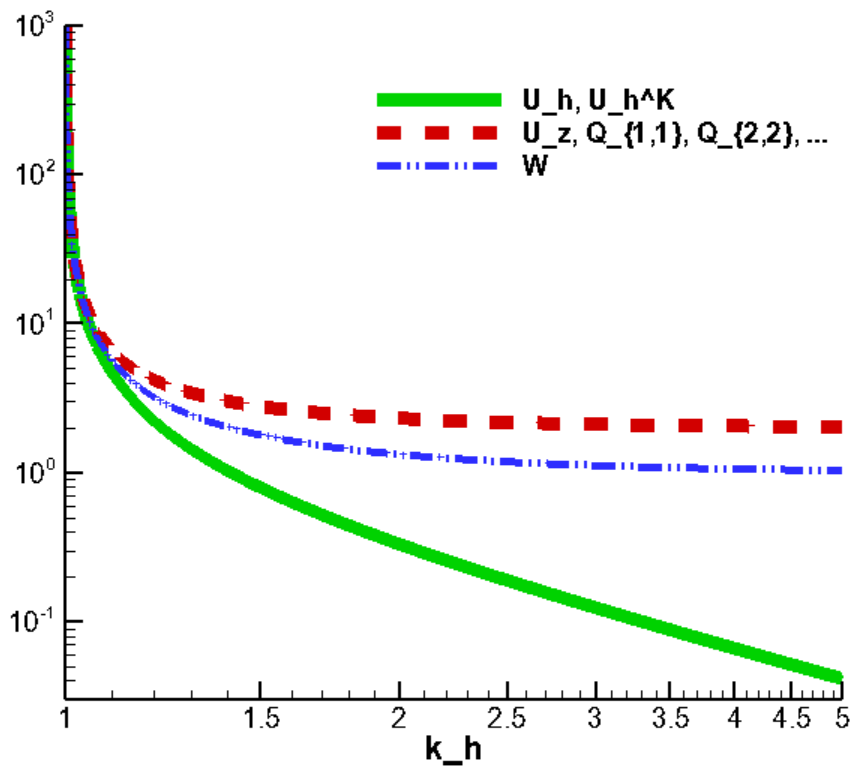}\\
  \caption{Sample modal spectra concentrating at the gravest modes ($\min\{k_h\} = 1$): $U^K_h$ (solid green) and $W$ (blue dashed dot dot) are plotted with $\Gamma_{\mathcal{E}}^K=-1$ and $\Gamma_{\mathcal{W}}^K=1.001$, which reproduces $U_h$ (the same solid green line) and $U_z$ (red dashed) in Fig. 2 of Zhu\cite{ZhuPoF18Schur} with $\Gamma_{\mathcal{E}}=-1$, $\Gamma_{\mathcal{W}}=2.001$, $\Gamma_{\mathcal{C}}=-1$ and $\Gamma_{\mathcal{Z}}=1$, and also others, such as $Q_{1,1}$, $Q_{2,2}$ et al. for the multi-scalar case with the corresponding parameter transformation indicated in the text.}\label{fig:2DdCae}
\end{figure}

Kraichnan\cite{K67} considered only the 2D flow and obtained
\begin{equation}\label{eq:K67}
U_h^K = (\Gamma_{\mathcal{E}}^K+\Gamma_{\mathcal{W}}^K k_h^2)^{-1}.
\end{equation}
Since the scalar is passive with no back reaction onto $\bm{u}_h$, so one, such as this author, might claim on the first sight that our result be pathological with $\Gamma_{\mathcal{Z}}$ and $\Gamma_{\mathcal{C}}$ ``wrongly'' entering the above $D$.\cite{ZhuPoF18Schur} But, actually,
\begin{equation}\label{eq:2Deq2D3C}
  \text{ we have $U_h^K = U_h$, with $\Gamma_{\mathcal{E}}^K=\Gamma_{\mathcal{E}}$ and $\Gamma_{\mathcal{W}}^K=\Gamma_{\mathcal{W}}-\Gamma_{\mathcal{C}}^2/\Gamma_{\mathcal{Z}}$,}
\end{equation}
which means nothing but a change of temperature parameters, instead of any pathology.
All the discussions of Kraichnan,\cite{K67} such as the negative-temperature state with large-scale concentration of energy and the implication of dual cascades, carry over to our case. For instance, according to Eq. (\ref{eq:2Deq2D3C}), we plot in Fig. \ref{fig:2DdCae} the graph of $U_h^K$ (and others for the multi-scalar cases) with $\Gamma_{\mathcal{E}}^K=-1$ and $\Gamma_{\mathcal{W}}^K=1.001$, which reproduces Fig. 2 of Zhu\cite{ZhuPoF18Schur} with $\Gamma_{\mathcal{E}}=-1$, $\Gamma_{\mathcal{W}}=2.001$, $\Gamma_{\mathcal{C}}=-1$ and $\Gamma_{\mathcal{Z}}=1$.
Of course, the passive scalar brings extra interesting phenomena and physical subtleties into the problem, and the discussions can be extended to multiple passive scalars advected by the same flow.


For $d=5$, however, the generalised helicity $\mathscr{S}$ from Eq. (\ref{eq:dDinvariantsD}) is now cubic. We can check that $\mathscr{S}$, unlike the Burgers situation explained in Appendix \ref{apd:ruggedness}, is not rugged to survive from the Galerkin truncation for constraining the absolute equilibrium whose spectral equipartition would then present a $k^4$ 1D spectrum. [Turbulence in $\mathbb{T}^5$ has already been simulated, though no (absolute eqillibrium) spectral information is presented by Yamamoto et al.\cite{YamamotoEApre12}] Note that we can reduce the 5D5C flow to 4D5C, 3D5C and even 2D5C systems with sequential imposition of cylinder conditions. And, yet more passive scalar(s), from reduction of dimension $d>3$ or not, with different initial fields and pumping mechanisms (if exist) can be included.
Interestingly, on the one hand, mathematically, it appears not making much sense talking about multiple passive scalars with identical diffusion and pumping operators (because they are simply the same mathematical object); but, on the other hand, if two or more scalars are subjected to different damping and/or pumping operators, we should go to their ideal advection equations, if we want to obtain any insights from the latter, and treat them as separate different ones. To be more explicit, for the 2D5C system case, we have three passive scalars $\theta_i$ ($i=1$, 2, 3), and, accordingly the passive scalar energy $\mathcal{Z}_{i}=\langle \theta_i^2\rangle$ and helicity $\mathcal{C}_{i}=2\langle \zeta \theta_i \rangle$. Any pair of mutual-corrrelation $\mathcal{Z}_{i,j} = \mathcal{Z}_{j,i} = \langle \theta_i \theta_j \rangle$ is also a rugged invariant and should be respected in the corresponding absolute-equilibrium analysis. Thus, introducing further the `temperature' parameters $\Gamma_{\mathcal{Z}_{ij}}=\Gamma_{\mathcal{Z}_{ji}}$, we have the canonical (J) PDF
\begin{equation}\label{eq:2D5Cnew}
\sim \exp\Big{\{} - \big{[} \sum_{i,j = 1}^{3}\Gamma_{\mathcal{C}_i} \mathcal{C}_i + \Gamma_{\mathcal{Z}_{ij}} \mathcal{Z}_{ij}+ \Gamma_{\mathcal{E}} \mathcal{E} + \Gamma_{\mathcal{W}} \mathcal{W} \big{]} \Big{\}}.
\end{equation}

One may be concerned by the possible mutual-correlation between different passive scalars or their derivatives. However, we see again from Eq. (\ref{eq:4Dhelicity}) that reducing from 4D4C to 2D4C with cylinder condition $\partial_{x_3} = \partial_{x_4} = 0$ results in the reduction:
$$\tau \to u_{3,2} u_{4,1} - u_{3,1} u_{4,2}.$$
Thus, explicitly, again the integration by parts show the vanishing mutual-correlation between the derivatives of the passively advected $u_3$ and $u_4$, reduced from the quadratic $\mathscr{N}$, should not affect the above result.

So, corresponding to Eqs. (\ref{eq:2D3CU} and \ref{eq:2D5Cnew}), we now have the 2D4C absolute equilibrium spectra
\begin{eqnarray}
U_h = \frac{\Gamma _{\mathcal{Z}_{11}}\Gamma _{\mathcal{Z}_{22}} - \Gamma _{\mathcal{Z}_{12}}^2}
{D_{24}}, \label{eq:2d4cUh}\\
Q_{C_1} \triangleq \langle \hat{\theta}_1\hat{\zeta}^* \rangle + c.c.= \frac{ -2 (\Gamma _{\mathcal{C}_1}\Gamma _{\mathcal{Z}_{22}}-\Gamma _{\mathcal{C}_2}\Gamma _{\mathcal{Z}_{12}}) k_h^2 }
{D_{24}} \ \text{and similarly $Q_{C_2}$}, \label{eq:2d4cQC1}\\
Q_{1,2} \triangleq \langle \hat{\theta}_1\hat{\theta}_2^* \rangle + c.c.= \frac{ 2(\Gamma _{\mathcal{Z}_{12}}\Gamma _{\mathcal{W}} - \Gamma _{\mathcal{C}_1}\Gamma _{\mathcal{C}_2}) k_h^2  +\Gamma _{\mathcal{Z}_{12}}\Gamma _{\mathcal{E}}}
{D_{24}}, \label{eq:2d4cQ12}\\
Q_{1,1} \triangleq \langle \hat{\theta}_1\hat{\theta}_1^* \rangle = \frac{ k_h^2(\Gamma^2 _{\mathcal{C}_{2}} - \Gamma _{\mathcal{W}}\Gamma _{\mathcal{Z}_{22}})  - \Gamma _{\mathcal{Z}_{22}}\Gamma _{\mathcal{E}}}
{D_{24}} \ \text{and similarly $Q_{2,2}$}, \label{eq:2d4cQ11}
\end{eqnarray}
with $D_{24} = \Gamma _{\mathcal{Z}_{11}}\Gamma _{\mathcal{Z}_{22}}\Gamma _{\mathcal{E}}-\Gamma _{\mathcal{Z}_{12}}^2\Gamma _{\mathcal{E}} + (\Gamma _{\mathcal{Z}_{11}}\Gamma _{\mathcal{Z}_{22}} \Gamma _{\mathcal{W}}-{\Gamma^{2}_{\mathcal{C}_2}}\Gamma _{\mathcal{Z}_{11}}-{\Gamma^{2}_{\mathcal{C}_1}}\Gamma _{\mathcal{Z}_{22}}+2{\Gamma_{\mathcal{C}_1}}{\Gamma_{\mathcal{C}_2}}{\Gamma _{\mathcal{Z}_{12}}}-\Gamma _{\mathcal{Z}_{12}}^2\Gamma _{\mathcal{W}} ) {k_h}^{2}$,
which can also be transformed to the Kraichnan spectrum with parameter transformations like Eq. (\ref{eq:2Deq2D3C}); thus $Q_{i,i}$ are also shown in the legend of Fig. \ref{fig:2DdCae}. 

One might also suspect that higher-dimensional invariants could have impact from dimensional reduction, which is not necessary in the current consideration because, for the problem of dimension $d>4$, the invariants of Eqs. (\ref{eq:dDinvariantsT},\ref{eq:dDinvariantsD}) are not quadratic and we would not expect the antisymmetry property as exposed for Burgers in Appendix \ref{apd:ruggedness}, thus not rugged. These $d$D ideal invariants can not be included in the absolute equilibrium analyses in the above, thus leaving no constraint on the corresponding statistical study of the reduced 2D or 3D passive scalars. The implication is that, for all $d>3$, the energy of the flow and that of the passive scalar are both expected to have absolute-equilibrium equipartition and turbulent forward transfer.

Some more remarks about the 3D problem are in order. As said, there are no further constraints with correlation between the flow and the passive scalars, but just as in 2D, we have the invariant pairing between the passive scalars themselves constraining the absolute equilibrium ensemble of the 3D multiple-scalar system. The absolute equilibrium (J)PDFs,
$\sim \exp \{ -[\Gamma_{\mathcal{E}}\mathcal{E} + \Gamma_{\mathcal{H}}\mathcal{H} + \sum_{i,j} \Gamma_{\mathcal{Z}_{ij}} \mathcal{Z}_{ij}] \}$,
for the flows and the advected passive scalars are simply the familiar independent Gaussian distributions, for each of the Fourier modes. We can easily obtain various statistical quantities. For instance, correpsonding to Eq. (\ref{eq:2d4cQ12}), now
$\{Q^{3D}_{i,j}\}$, which is the matrix inverse $\{ \Gamma_{\mathcal{Z}_{ij}} \}^{-1}$, is independent of $\bm{k}$ --- equipartitioned. 
And, of course, the `coherency spectra' matrix $\{Q^{3D}_{i,j}(Q^{3D}_{i,i}Q^{3D}_{j,j})^{-1/2} \}$is also equipartitioned. As discussed in the end of the last subsection, `bottleneck' and `reduction of intermittency growth' is then also expected for the accordingly related quantities.
Passive scalar turbulence with diffusion and dissipation are of course far from the absolute equilibrium, but the importance of the quadratic interaction responsible for the thermalization still indicates that the initial linear dependence, if any, between 3D passive scalars and the flow tend to be de-correlated (not by the damping of the amplitudes, because we are talking about normalized statistical correlation coefficient) at least in the homogeneous setup, which can be checked in physical or \textit{in silico} experiments (c.f., e.g., Fig. 15 in Ref. \onlinecite{sirivatWarhaftJFM82}) and which should be respected in modeling. Similar results also hold for $d > 3$, though for the different reason concerning ruggedness as mentioned in the last sub-section, with the absolute equilibrium (J)PDFs
$\sim \exp \{ -[\Gamma_{\mathcal{E}}\mathcal{E} + \sum_{i,j} \Gamma_{\mathcal{Z}_{ij}} \mathcal{Z}_{ij}] \}$.

The independent white-noise velocity assumption in the Kraichnan\cite{K68} model thus appears to make a bit more sense in the spaces of $d>2$. Note however that systematic perturbative calculations\cite{FGVrmp01} show an intermittency correction to the normal inertial scaling exponent (other than the second-order one). The correction decreases with $d$, and vanishes as $d \to \infty$, which however we can not connect with the current absolute equilibrium analysis.


\section{Further discussions}
We have made very basic absolute equilibrium analysis of the high dimensional flows and their cylindrical reductions resulting in passive scalars. Several physical conjectures and explanations, relevant to numerical simulations and experiments (e.g., Refs. \onlinecite{SuzukiEApof05,GotohETC4DNS07,sirivatWarhaftJFM82}) have been offered accordingly. The most impressive result probably is that in 4D and its reduction to 3D the quadratic ideal invariant related to the vorticity function\cite{ArnoldKhesin98Book} happens to vanish without boundary contribution, and that the absolute equilibrium scalar is equipartitioned and not correlated to the flow; while, 2D passive scalar is constrained by the cylindrical reduction of the 3D helicity which correlates it to the flow and leads to non-equipartition.
One should also be careful with the statement in Zhu\cite{ZhuPoF18Schur} that the relaxation toward the helical absolute equilibrium with $\mathcal{Z}$ concentrated at large scales should not induce inverse cascade/transfer just among the $k_z = 0$ modes, which we now feel not so appropriate though it is possible that such an internal channel may be very sensitive and unstable to perturbations (such as the $u_z$ wave remarked in Ref. \onlinecite{ZhuPoF18Schur}) and can yield to other external routes.

We iterate that $\mathcal{C}$ is formed only when a passive scalar $\theta$ in the incompressible two-dimensional flow $\bm{v}$ is so smart as to track the (signed) intensity of the vorticity field, rather than $\bm{v}$ tracking $\theta$ (which would imply that $\theta$ had back reaction onto $\bm{v}$). This means that in an experiment (physical or \textit{in silico}) of two dimensional flows, the tracer should be put into vortical regions of the flow in a selective way, or that the passive scalar is somehow by itself smart
enough, at least in the pumping scales [the wavenumber(s) where $\mathcal{C}$ is injected in a numerical simulation, say] when entering the system, to locate itself in a right coherent way. Such scalar is still quite different to the vorticity which may present ``ideal'' turbulence in the sense of zero molecular diffusion limit\cite{Onsager49, Frisch95book} with forward $\mathcal{W}$ cascade of $k^{-1}$ spectrum (with logarithmic correction, according to Kraichnan\cite{K67}), thus $\mathcal{W} \to \infty$ and not incompatible with current rigorous mathematical results of 2D Euler;\cite{EyinkPhD08} and, even though $\mathcal{Z}$ can cascade forwardly like $\mathcal{W}$, genuine (simultaneous or not) ``leaking''/inverse transfer to large scales of a fraction of it can not be excluded so far.

The last remark in the above is relevant, because, if we suppose in Eqs. (\ref{eq:p1}) and (\ref{eq:p2})
\begin{equation}\label{eq:equalForcing}
\text{$f_{\theta} = f_{\zeta} = \nabla \times \bm{f}_{\bm{v}}\cdot \bm{z}$, then $(\partial_t + \bm{v} \cdot \nabla - \nu \nabla^2)  (\theta-\zeta) = 0$, given $\nu = \kappa$}.
\end{equation}
However, with finite $\mathcal{W}$, no dissipative anomaly is expected (c.f., Ref. \onlinecite{EyinkPhD08} and references therein) in the $\nu \to 0$ limit, thus no reason for asymptotic $\theta=\zeta$; with infinite $\mathcal{W}$, the assumed dissipative anomalies may not be the same (even with $\nu = \zeta$ vanishing equally) and dissipative anomalies would imply \textit{intrinsic stochasticity},\cite{CCMVnjp04,DrivasEyinkJFM17I} neither any reason for $\zeta = \theta$. It appears precisely because $\theta$ is (strongly) correlated to $\zeta$ thus associated to the inversely cascaded $\bm{v}$ while cascading forwardly with $\zeta$, but without the relation $\theta = \zeta$, so that some $\mathcal{Z}$ can `leak' along larger and larger scales. Such a $\nu = \kappa$ equally-vanishing asymptotic scenario indicates that we need very small diffusivity (and accordingly very high resolution in numerical simulations) and would probably need to wait a long time, for the nonlinear dynamics to beat the molecular linear damping (and probably other numerical dissipation effect etc.), to see the inverse transfer and large-scale concentration of $\mathcal{Z}$ (if indeed). A large-scale `friction' or, in terms of the numerical experiments, the \textit{hypo-diffusivity} with increasing damping at larger scales could help facilitating the inverse $\mathcal{Z}$ transfer. In rotating flows at intermediate/moderate Rossby numbers presenting large-scale accumulation of vertically-averaged vertical velocity energy,\cite{CCEH05} it may happen that ``$u_z$-waves deposit energy into large-scale $u_z$-vortex but then extract basically the same amount from other vortex modes (of $u_z$ at small scales or, so far not excludable, of $\bm{u}_h$ at any scales)'',\cite{ZhuPoF18Schur} but the pure 2D passive-scalar mechanism described above may also contribute, directly or indirectly.

Recently, Linkmann et al.\cite{LinkmannETAepje18} performed very interesting direct numerical simulations of helical 2D3C flows. They claimed, by superficially the same fashion of the equations for 2D passive scalar $\theta$ and 2D vorticity $\zeta$, that $\theta$ and $\zeta$ should converge to each other as time goes, contrary to what we argued in the beginning of the last paragraph (actually several years ago\cite{Zhu2D3Cold}). They touched, though not quite confirmed, the scenario of non-universal transfer of the $\theta$ energy anticipated in Ref. \onlinecite{Zhu2D3Cold}. However, that paper does not appear to be conclusive. 
They presented the convergence of $\theta$ and $\zeta$ spectra to each other at smaller wavenumbers than the forced one(s) $k_f$, presumably wanting the damping mechanism for the argument they proposed. And, large-scale damping, say, the hypo-diffusivity/viscosity, was not implemented to sustain a stationary state. Their late-time result with no inverse flux of $\mathcal{Z}$ may be due to the non-local interactions between modes of $k<k_f$ and $k>k_f$, the latter, strongly damped, taking some energy from the former to cancel the inverse flux due to other interactions. Also, we take issue with their statement that $u_z=\theta$ would turn into an active scalar in helically forced 2D3C flows. In fact, the forcing $\bf{f}_{\bm{v}}$ or $f_{\zeta}$, if exists, on the 2D advecting flow does not need to be a function of $\theta$ to have effective helicity injection rate $2\langle \zeta f_{\theta} + \theta f_{\zeta}\rangle$: We can always appropriately choose $f_{\theta}$ to effectively inject helicity, for whatever $f_{\zeta}$ chosen to be  $\theta$-independent for a passive scalar problem. 
More systematic numerical and experimental examinations are still wanted. 

Finally, with the notion of semidirect product (e.g., Ref. \onlinecite{MarsdenRatiuWeinstienAMS84} and references therein) and dealing with the (Lie algebra of) operator of the sum of Lie derivative and density $L_{\bm{v}}+\rho$, Khesin and Chekanov\cite{KhesinChekanovPhD89} have generalised the integrals (\ref{eq:dDinvariantsT},\ref{eq:dDinvariantsD}) to more general barotropic flows. We see that in general the varying density $\rho$ is involved in the continuum of the Casimir functions $\mathscr{T}=\int_{\mathbb{T}^4} g(\tau)\rho\mu$ with $\tau = (\verb"d" \verb"U")^m/(\rho \mu)$ in Eq. (\ref{eq:dDinvariantsT}), but it is not the case with $g = I$ where $\rho$ cancels out:
Now, formally the same as in the incompressible case, $\mathscr{T}= \mathscr{N}$, with $m=2$ for 4D flows in the `nontrivial' quadratic integral, thus vanishing, and thus `trivially nontrivial'.  Interestingly, then Kraichnan's\cite{K55} analysis respecting only energy would be more favorable in dimension $d>3$, unlike in 3D turbulence where the helicity integral can polarize the energy partitions.\cite{ZhuJFM16}

\begin{acknowledgements}

This work is supported by NSFC (No. 11672102) and  Ti\'an-Yu\'an-Xu\'e-P\`ai (No. 27182818) fundings.

\end{acknowledgements}

\appendix

\section{Cylinder from thinning}\label{apd:cylinder}
\begin{figure}
  \centering
  \includegraphics[width=
  \paperwidth
  ]{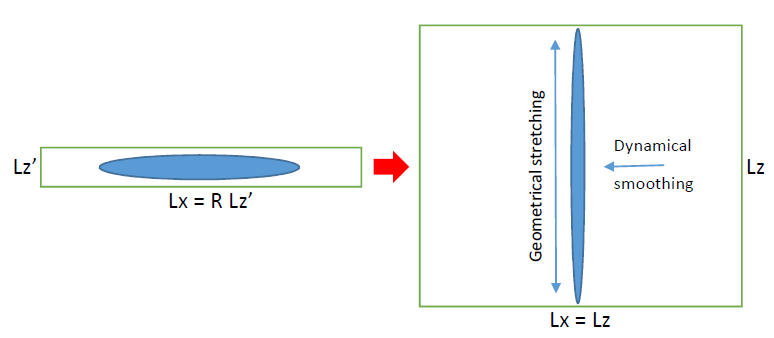}\\
  \caption{2D sketch of the transform to approach the cylindrical reduction: the shaded structure is not only geometrically stretched but also dynamically smoothed more in the $z$ direction (than in the $x$ direction) to have asymptotically $\partial_z \to 0$ almost everywhere.}\label{fig:transform2cylinder}
\end{figure}

For simplicity, we start with the three-dimensional (3D) incompressible Navier-Stokes (NS) equation, (\ref{eq:p3}) below, in a box thinning by a factor $R$ in the $z'$ dimension (e.g., Ref. \onlinecite{CMVprl10}), i.e.,
\begin{equation}\label{eq:2D3Cscaling1}
RL_{z'} = L_x=L_y=L
\end{equation}
(as scketched by Fig. \ref{fig:transform2cylinder} in the $x$-$z$ plane).
It turns out that such thinning is effectively introducing the (asymptotic) cylinder condition, because we have the re-scaled anisotropic viscosities
\begin{equation}\label{eq:2D3Cscaling2}
\text{$R^{-2} \nu_{z} = \nu_x=\nu_y=\nu$, with $z=z'R$}
\end{equation}
\text{with $z=z'R$}
and the corresponding re-scaling of vertical (along-$z'$) velocity and forcing (pressure can be canceled by incompressibility.)
The resultant much larger $\nu_{z}$ with large $R$ smoothes the $z$ dynamics more, indicating much smaller $\partial_{z}$, formally $\to 0$ (`cylinder condition'), i.e., asymptotically the two-dimensional-three-component (2D3C) state, with $R\to \infty$ and the vertical velocity $u_{z}$ passively advected by the 2D `horizontal' flow, thus a 2D passive scalar problem with unit Schmit/Prandtl number.

There are also other dynamical processes, such as rapid rotating flows (c.f., for example, Bourouiba\cite{BourouibaJFM12} and Mininni et al.\cite{MininniETApre11} who also studied relevant absolute equilibria), that lead to (partial) two-dimensionalization.
\section{Rugged invariant: heuristic argument with the caveat rigorously illustrated by the Burgers dynamics}\label{apd:ruggedness}
\begin{figure}
  \centering
  \includegraphics[width=\paperwidth]{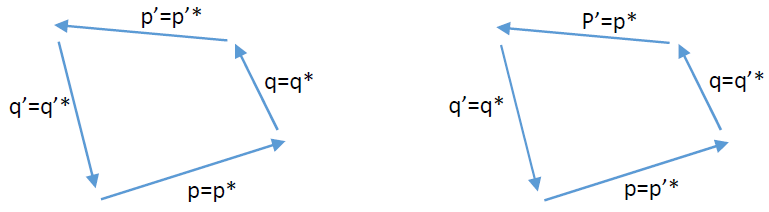}\\
  \caption{For better visualization, we pretend to work in the 2D plane (otherwise, all wave vectors will fall onto a line, hard to be distinguished with bare eyes in the figure.) For each set of modes with $p=p*$, $q=q*$, $p'=p'*$ and $q'=q'*$, there are corresponding set of modes with $p=p'^*$, $q=q'^*$, $p'=p^*$ and $q'=q^*$ in Eq. (\ref{eq:caveat}) according to which $p^*+q^* = -(p'^* + q'^*)$, thus the contributions of this, and all, pairs of sets of modes cancel exactly.}\label{fig:nullBurgers}
\end{figure}

Consider for simplicity the 1D dynamical variable $v$ in a torus of period $2\pi$. Suppose, for instance, we have a conservation law
\begin{equation}\label{eq:heuristic1}
\frac{d\langle v^3 \rangle}{dt} = 0,
\end{equation}
with
\begin{equation}\label{eq:heuristic2}
\langle v^3 \rangle \triangleq \int_0^{2\pi}[v(r,t)]^3 dr/2\pi = \sum_{k+p+q = 0} \hat{v}_k \hat{v}_p \hat{v}_q.
\end{equation}
Now define the Galerkin truncation
\begin{equation}\label{eq:heuristic3}
\text{$v \to \bar{v}\triangleq P_G v = \sum_{|k|<K} \hat{v}_k \exp\{-\hat{i}kr \}$ with $\hat{i}^2 = -1$ and $\hat{v}_k = 0$ for $|k|>K$},
\end{equation}
i.e., with modes of their wavenumber modules greater than $K$ being removed.
Note however that, due to nonlinear interaction/mode coupling,
\begin{equation}\label{eq:heuristic4}
\text{for $2K\ge |k| > K$, $\frac{d\hat{v}_k}{dt} \ne 0$}
\end{equation}
in general. Then, we have
\begin{equation}\label{eq:heuristic5}
\frac{d\langle \overline{ \bar{v}^3 } \rangle}{dt} = \frac{d\langle v^3 \rangle}{dt} - \sum_{k + p + q = 0}^{|k| > K, |p| \le K, |q| \le K} \frac{d\hat{v}_k}{dt}\hat{v}_p \hat{v}_q 
\end{equation}
which does not necessarily vanish. In other words, such an invariant does not survive after the Galerkin truncation, i.e., not rugged, and this should be generic for other non-quadratic invariants, if any, for our $d$D incompressible flows. The caveat in the above analysis is that specific dynamics have not been further exploited. In fact, the Burgers equation,
\begin{equation}\label{eq:Burgers}
\frac{d\hat{v}_k}{dt} = -\frac{\hat{i}}{2}\sum_{p'+q'=k}  k \hat{v}_{p'}\hat{v}_{q'},
\end{equation}
is an exception, for which, with all $|p|,|p'|,|q|,|q'| \le K$,
\begin{equation}\label{eq:caveat}
\sum_{k = -( p + q )}^{|k| > K,
} \frac{d\hat{v}_k}{dt}\hat{v}_p \hat{v}_q = -\frac{\hat{i}}{2} \sum_{p' + q' =-( p + q )}^{|p' + q'| > K} (p' + q') \hat{v}_{p'} \hat{v}_{q'}\hat{v}_p \hat{v}_q  
\end{equation}
in Eq. (\ref{eq:heuristic5}) indeed is seen to vanish with the antisymmetry property ({\it c.f.}, Fig. \ref{fig:nullBurgers}) in the above right-hand side. Thus, the Hamiltonian $\langle v^3 \rangle$ is proved, in an explicit way other than Abramov et al.,\cite{AbramovKovacicMajdaCPAM03} to be a rugged invariant.  For $\langle v^4 \rangle$, five modes with $p'+q'=-(p+q+r)$ without the antisymmetric property, instead of the four-mode $p'+q'=-(p+q)$ with antisymmetry in Eq. (\ref{eq:caveat}), are involved, besides other subtleties, and we do not have the same antisymmetry property to be used to prove the ruggedness; similarly for other higher powers. Another counter example with rugged non-quadratic Hamiltonian is the nonlinear Shr\"{o}dinger  (e.g., Ref. \onlinecite{KrstulovicBrachet11} with cubic nonlinearity). Similarly bringing the dynamics into the calculations, it is possible to argue more systematically that other non-quadratic invariants of $d$D ideal Euler hydrodynamics are unrugged. For example, for $g(v)=v^3$ Eq. (\ref{eq:dDinvariantsT}) with $d=2m$, although the first term in the right hand side of Eq. (\ref{eq:heuristic5}) vanishes, the value of the term behind it depends on the solution, for the transverse projection of the pressure destroys the antisymmetry property that appears in the Burgers (a systematic proof for all other powers however are still wanted). As remarked in the introductory discussions, we iterate that, even though there are other schemes, say, relevant to the $sine$-algebras, of finite-degree approximations resulting in more (analogous) invariants, hardly exists any analytical expression of useful observable, the energy spectrum, say, with direct physical illuminations; and that, even for other models providing the caveats, though nicely documented and analyzed, effective physical predictions are not clearly available from Abramov et al.\cite{AbramovKovacicMajdaCPAM03}, Krstulovic and Brachet\cite{KrstulovicBrachet11} or Krstulovic et al.\cite{KrstulovicETC09}

If an invariant is quadratic, say $\langle v^2 \rangle$, for which the equation corresponding to and in contrast to (\ref{eq:heuristic5}) is
\begin{equation}\label{eq:heuristic6}
\frac{d\langle \overline{ \bar{v}^2 } \rangle}{dt} = \frac{d\langle v^2 \rangle}{dt} - 2 \sum_{|k| > K} \frac{d|\hat{v}_k|}{dt}|\hat{v}_k| = 0,
\end{equation}
then we see that we have proved the ruggedness of the known ideal hydrodynamic quadratic invariants, which can also be directly checked\cite{Onsager49,K73} from the Galerkin truncation Euler dynamics
\begin{equation}\label{eq:PGdynamics}
\partial_t \bar{\bm{v}} + \overline{\bar{\bm{v}}\cdot \nabla \bar{\bm{v}}+ \nabla P} = 0.
\end{equation}
While for the rugged invariant surviving from the truncation, we don't even bother introducing the extra (over)bar to distinguish the notations. Note that the above Eq. (\ref{eq:PGdynamics}) is defined only for $|k| \le K$, thus Eq. (\ref{eq:heuristic4}) is allowed.

\end{document}